\documentclass[conference]{IEEEtran}

\usepackage{svg}
\usepackage{multirow}
\usepackage{amsmath}
\usepackage{float}
\usepackage{subcaption}
\usepackage{graphicx}
\usepackage[utf8]{inputenc} 
\usepackage[T1]{fontenc}    
\usepackage{hyperref}       
\usepackage{url}            
\usepackage{booktabs}       
\usepackage{amsfonts}       
\usepackage{nicefrac}       
\usepackage{enumitem}

\title{Optimizing an RFSoC LLRF Controller for a Low-PRF, Short Pulse System}

\author{
    \IEEEauthorblockN{
        Victoria Moore\IEEEauthorrefmark{1},
        Qiang Du,
        Shreeharshini Murthy,
        Michael Chin,
        Angel Jurado,
        Benjamin Flugstad   
    }
    \IEEEauthorblockA{
        Lawrence Berkeley National Laboratory\\
        1 Cyclotron Rd, Berkeley, CA, 94720 USA
    }
}

\begin{document}
\maketitle

\begin{abstract}
    The Advanced Light Source (ALS) LINAC operates with 3 GHz RF pulses at 0.7 Hz pulse repetition frequency (PRF) and 1.2 µs duration, imposing severe limitations on traditional PID and adaptive feedforward control schemes. An RFSoC-based controller prototype was developed and presented in prior work to upgrade the legacy LLRF systems for the LINAC and Subharmonic Buncher.
This paper presents sampling scheme optimizations and interleaving calibration techniques for RFSoC based control of pulsed RF systems.
We detail bench characterization methodologies to isolate component-level phase noise across individual ADC, DAC, and clock components in the Xilinx ZCU208 hardware ecosystem.
Finally, we explore translating open-loop controller bench measurements into controller-added closed-loop LLRF performance predictions.
\end{abstract}

\section{Introduction}
ALS is currently upgrading the LINAC and Subharmonic Buncher Low Level RF control systems. The RFSoC-based platform was described in previous work \cite{du:2025} and the PYNQ-based firmware design is available on GitHub \cite{lbnl:pynq_llrf}. Initial RF characterizations and clocking optimizations indicated performance comparable to other digital LLRF platforms for normally conducting cavities \cite{shree:2025}. 
Recent work has focused on developing measurement metrics to evaluate if the RFSoC-based platform can achieve the performance required for the LINAC application. 

\subsection{ALS LINAC System}
The two 3 GHz LINAC RF stations are pulsed at 0.7 Hz, 1.2$\mu$s pulses. The 400W pre-amplifier and the 17 MW klystron are currently run in saturation with an open loop legacy LLRF analog amplitude and phase control. The very low Pulse Repetition Frequency (PRF) caused the Thales klystrons to arc during initial commissioning. System optimizations required using the cathode pulse end to control the tail of the RF pulse to widen the pulse length without arcing. Future upgrades may include increasing the cathode PRF to reduce arcing and adding a downstream SLAC Energy Doubler (SLED) to widen the RF pulse. 

The RFSoC controller prototype was used to characterize the existing RF waveforms in the Accelerating Sections (AS) cavities 1 and 2.
\begin{figure}
    \centering
    \includegraphics[width=1\linewidth]{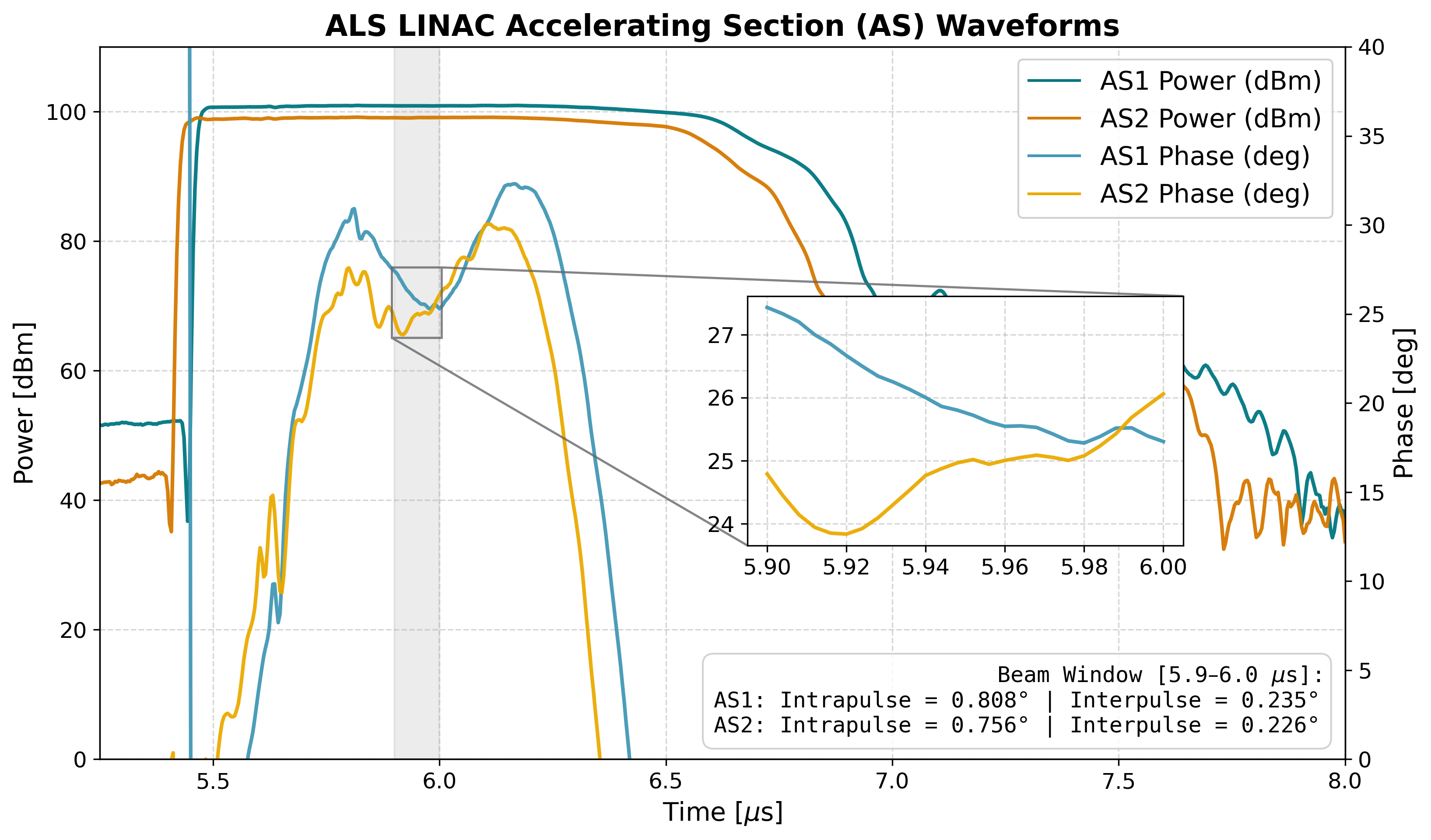}
    \caption{ALS LINAC Accelerating Sections 1 and 2 FWD and REV Waveforms}
    \label{fig:placeholder}
\end{figure}
From these initial measurements we observed a highly non-linear, but repetitive phase profile through the pulse, confirming the need for improved phase control. The new system will provide open loop drive and closed loop phase control to avoid interfering with the existing high power RF system tuning.
\subsection{LINAC LLRF Performance Requirements}
The project requested 0.1$^\circ$ RMS RF field stability in the 3 GHz normal conducting cavities. Future work will quantify open-loop performance of the high-power RF system. Due to the low PRF and short pulses, the control system's ability to correct for high-power RF errors is limited, so the LLRF platform inherent noise performance is critical. 
The system front end RF characterization showed a negligible affect on system linearity or stability \cite{du:2025}, so current effort is focused on characterizing the ZCU208 platform, RF-identical to the modified 'LBL208' platform that will be used in production. Among other minor optimizations, the LBL208 swaps the XCZU48DR-2FSVG1517 chip for the cheaper XCZU47DR-1FSVG1517 and is currently in production for the ALS-U BPM systems. \cite{Chin:MORT07}. 

\subsection{Project Status}
The team is manufacturing chassis hardware and implementing the required custom python software on top of the PYNQ framework and our custom IP design, utilizing our standard 250 MHz LLRF DSP blocks. Commissioning is planned before the upcoming ALS Dark Time for the ALS 2.0 Upgrade, October 2027. 
\section*{ADC Interleaving Spur Calibration}
\subsection*{Background}
The RFSoC achieves GS/s sampling rates by interleaving multiple ADCs into tiles. 
Imperfections and mismatches between the sub ADCs manifest as spurs at $k*F_s/N$, where $F_s$ is the composite sampling rate of the RF-ADC, N is the number of sub-RF-ADCs interleaved together, and $k = 0 \dots N-1$.
The RFSoC provides calibration algorithms to correct for DC offsets, gain differences, and sampling clock time skew between the sub ADCs.
\\
\indent
The foreground calibration runs as part of the RF-ADC initialization state machine sequence to correct the DC offsets, so there should be no signal energy present at $k*F_s/N$ during ADC startup. 
\\
\indent
After the RF-ADC initializes, the background calibration updates the corrections as the environment changes. The background calibration process requires an input signal of -40 dBFS, and blocks require 7 $\mu$s to change state. For our 1.2 $\mu$s pulse length application, we will calibrate the ADCs with a test signal and then freeze these calibration weights, disabling the background calibration. 

\subsection*{Interleaving Spur Locations}

\begin{table}[ht]
\centering
\begin{tabular}{l|r}
Imperfection & Spur Location \\\hline
DC offset & $k*F_s/N$ \\
Gain &  +/-$f_{in} + (k/N)*F_s$ \\
Time skew & +/-$f_{in} + (k/N)*F_s$

\end{tabular}
\caption{\label{tab:widgets}RFSoC Interleaving Spurs.}
\end{table}

Digital downconversion shifts the residual gain and timing skew spurs to baseband. 

In the RFSoC, the  FIR filter and the rate-reduction sample decimation are implemented as a polyphase filter which performs FIR filtering on the samples that will be preserved through decimation. 
This limits the unfiltered in-band content to $-\frac{f_s}{2M}$ to $\frac{f_s}{2M}$, and if $\frac{f_s}{2M} \le \left\vert{} k \cdot \frac{f_s}{N} \right\vert{}$, where M is the decimation factor, all of the gain and timing skew spurs will be filtered before decimations folds them into baseband. 

With the background calibration enabled, the timing and gain skew spurs were suppressed below the noise floor in our measurements. 

If the $f_{RF} = \frac{A*f_s}{2M}$, where A is an integer, then the filtered timing and gain skew spurs will fold directly onto DC in a zero-IF architecture, preventing the background calibration from detecting the skew between sub-adcs.
Uncalibrated, the filtered but uncalibrated skew spurs measured 60-80 dBc near 3 GHz (7.5-75 fs at 3 GHz).
In these schemes the background calibration can be performed with a test signal near $f_{RF}$.
With our 4 GS/s $f_{s}$ and 3 GHz $f_{RF}$ scheme, performing a background calibration with the NCO offset by a few MHz suppresses the gain and timing skew spurs below 90 dBc for nominal operation: a residual 1.7 fs phase error. Future work will move to a 5 GS/s $f_s$ to mitigate this issue. 

\section*{Characterization Measurements}
\subsection*{ADC 3 GHz Spectrum}
With the frozen, offset interleaving calibration in place, the ADC spectrum can be evaluated:
\begin{figure}
    \centering
    \includegraphics[width=1\linewidth]{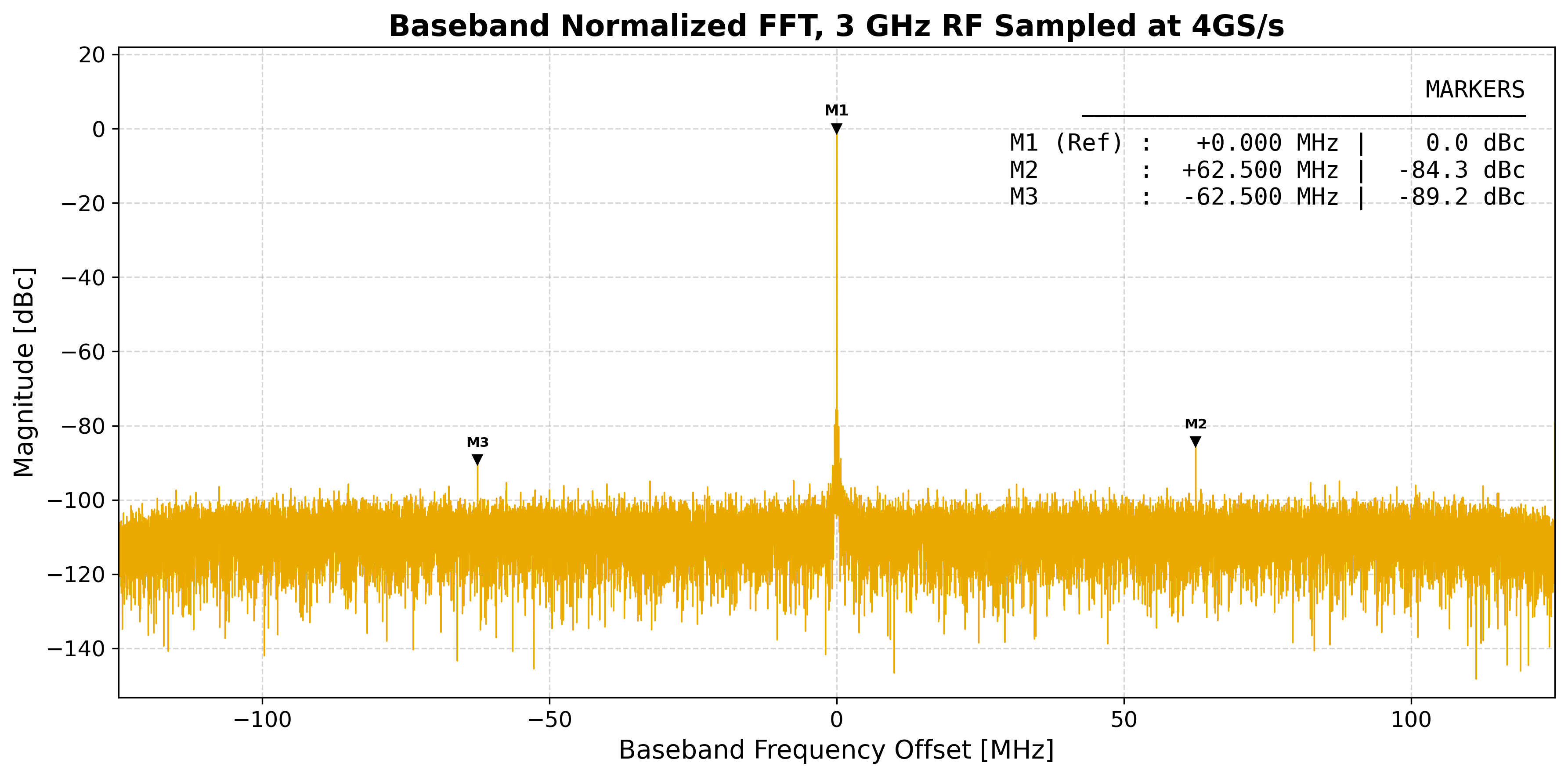}
    \caption{Baseband Spectrum, 3GHz $RF_{\text{IN}}$, 4 GS/s $f_s$}
    \label{fig:placeholder}
\end{figure}
The 62.5 MHz spurs will be investigated for wideband diagnostic performance improvement in future work.  
\subsection*{DAC 3 GHz Spectrum}
\begin{figure}[H]
    \centering
    \includegraphics[width=1\linewidth]{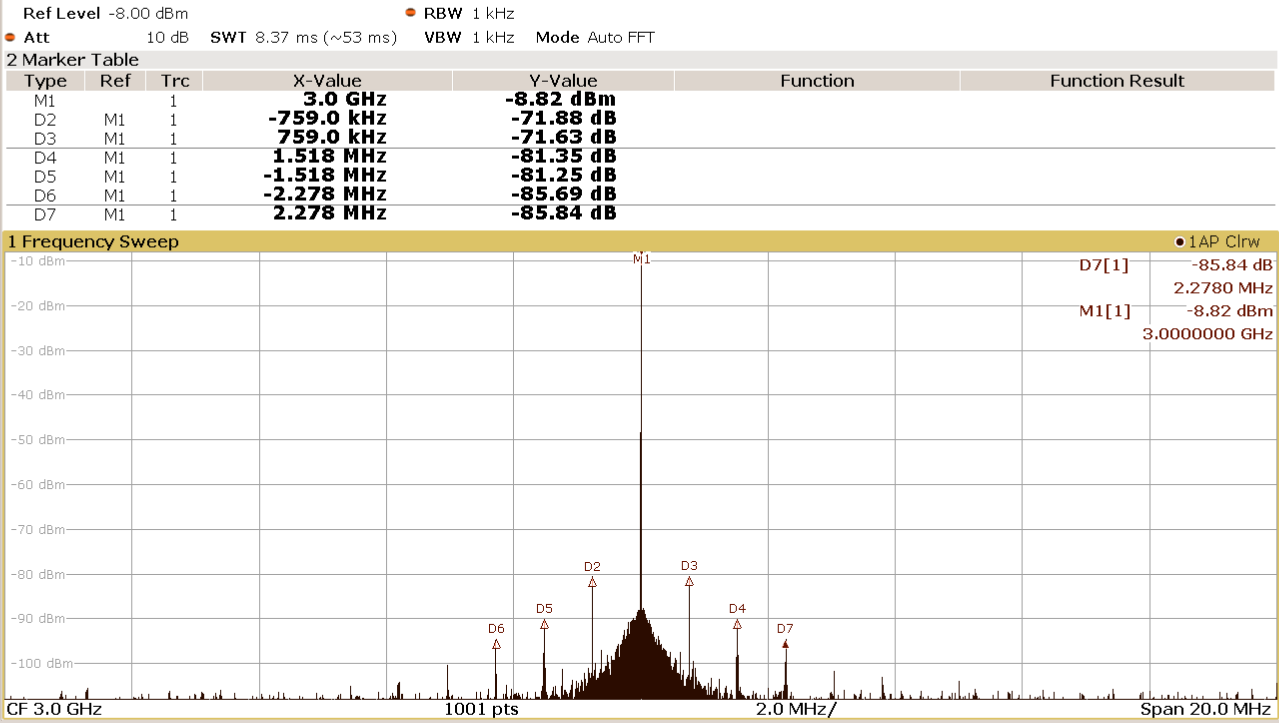}
    \caption{DAC Spectrum, 3GHz $RF_{\text{OUT}}$, 4 GS/s $f_s$, with 70 dBc  759 kHz modulating spurs}
    \label{fig:placeholder}
\end{figure}

\subsection*{Sampling Clock Phase Noise}
The Xilinx CLK104 board was modified to the 'CLK104A': adding 0$\Omega$ jumpers to connect the J11 'external clock input' connector to the LMK04828 to be used as a buffer and a divider for clock generation \cite{shree:2025}. For the 4 GS/s LLRF design, the LMX2594 synthesizer was optimized for low jitter. The ALS BPM team is also using the CLK104A, configured differently to allow the recovered clock from the timing link to be used as the time base reference \cite{Chin:MORT07}.
\begin{figure}[H]
    \centering
    \includegraphics[width=1\linewidth]{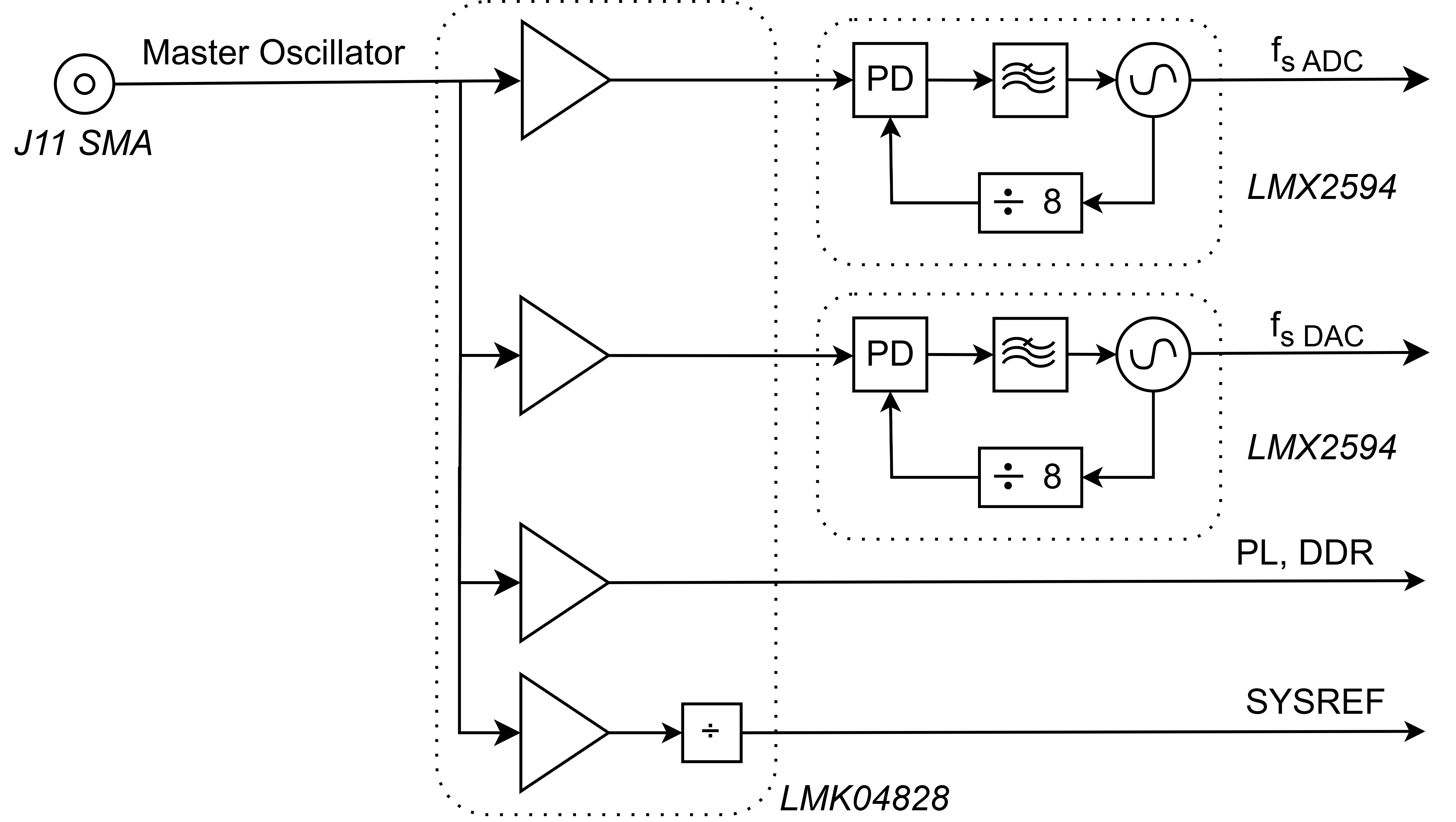}
    \caption{Simplified CLK104A Diagram}
    \label{fig:placeholder}
\end{figure}
The CLK104A and DAC phase noise measurements were made with the Rhode and Schwartz FSWP8 Phase Noise Analyzer, calibrated to achieve <1fs added jitter. The LMX1204 is a very low phase noise clock divider, <1fs as a 3 GHz buffer, which combined with the FSWP enables an additive phase noise measurements across frequency. 
\begin{figure} [H]
    \centering
    \includegraphics[width=1\linewidth]{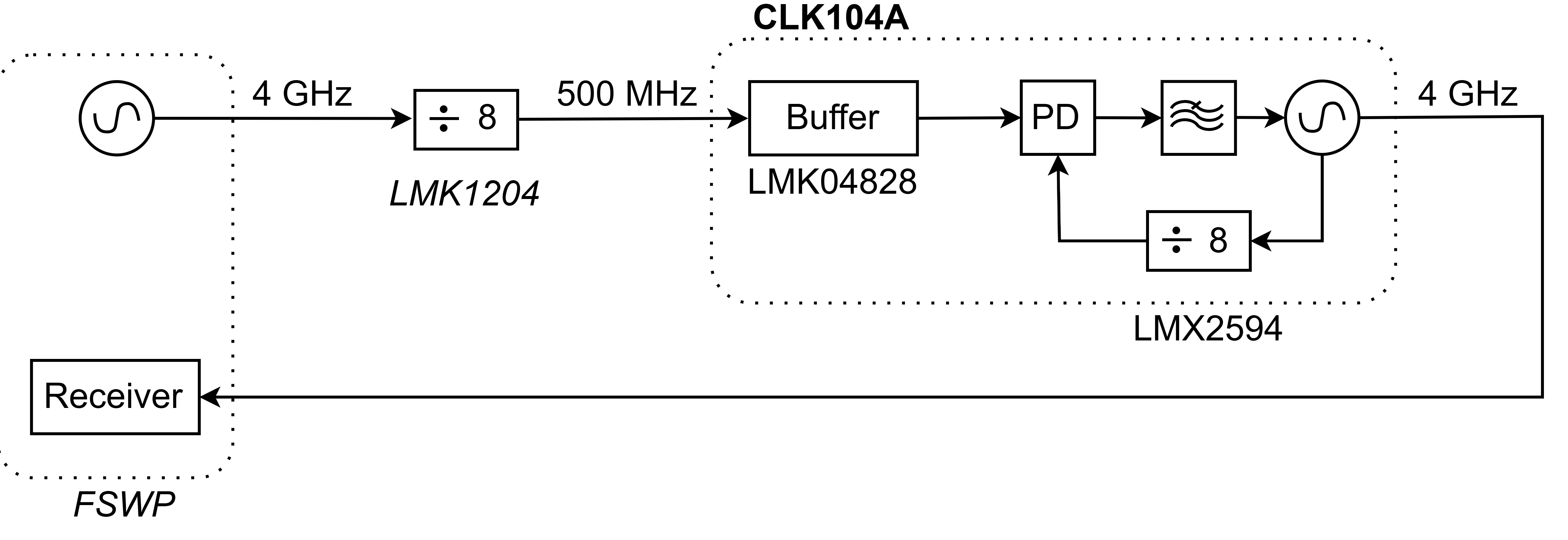}
    \caption{CLK104A Phase Noise Measurement Setup}
    \label{fig:placeholder}
\end{figure}
\begin{figure} [H]
    \centering
    \includegraphics[width=1\linewidth]{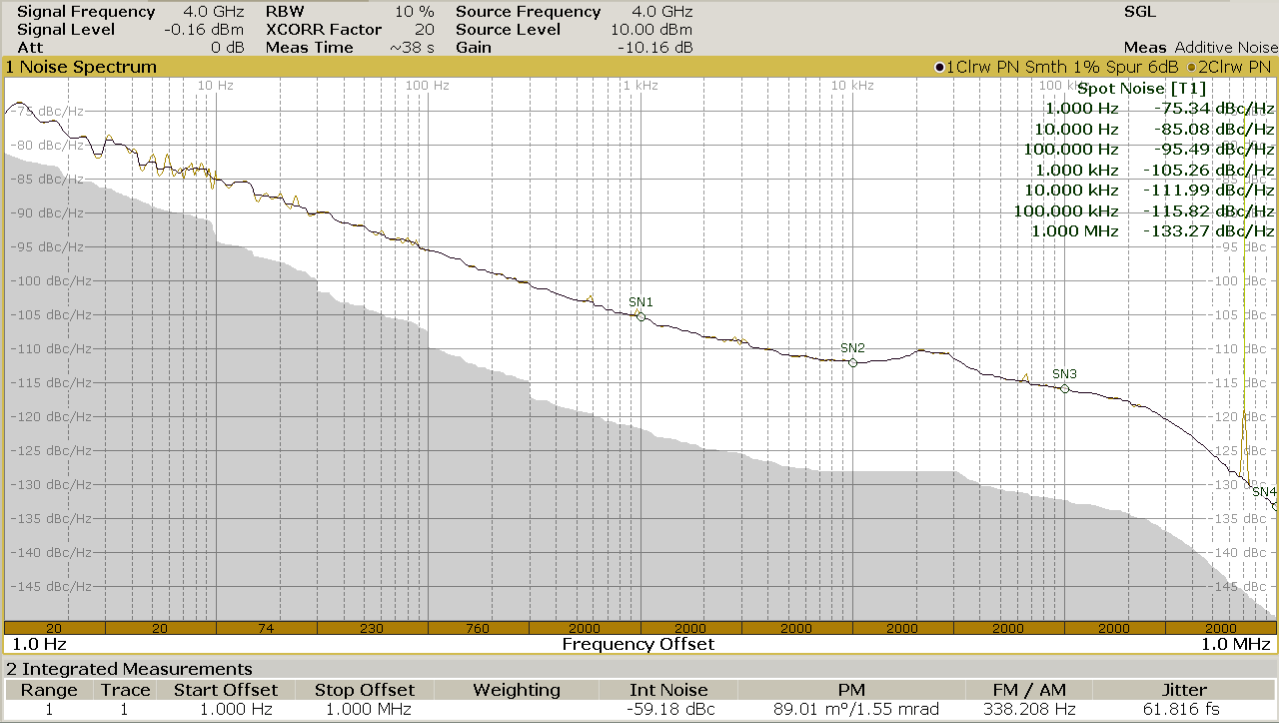}
    \caption{CLK104A Additive PN, 500MHz $RF_{\text{IN}}$, 4GS/s $f_s$} 61.8 fs [1 Hz, 1 MHz]
    \label{fig:placeholder}
\end{figure}

The CLK104A sampling clock generation impacts the phase noise performance but is required for the current ALS MO architecture, which distributes 500 MHz RF over coax to control stations. 

\subsection*{DAC Phase Stability}
Sampling clock generation dominates the DAC phase noise performance. 
\begin{figure}[H]
    \centering
    \includegraphics[width=1\linewidth]{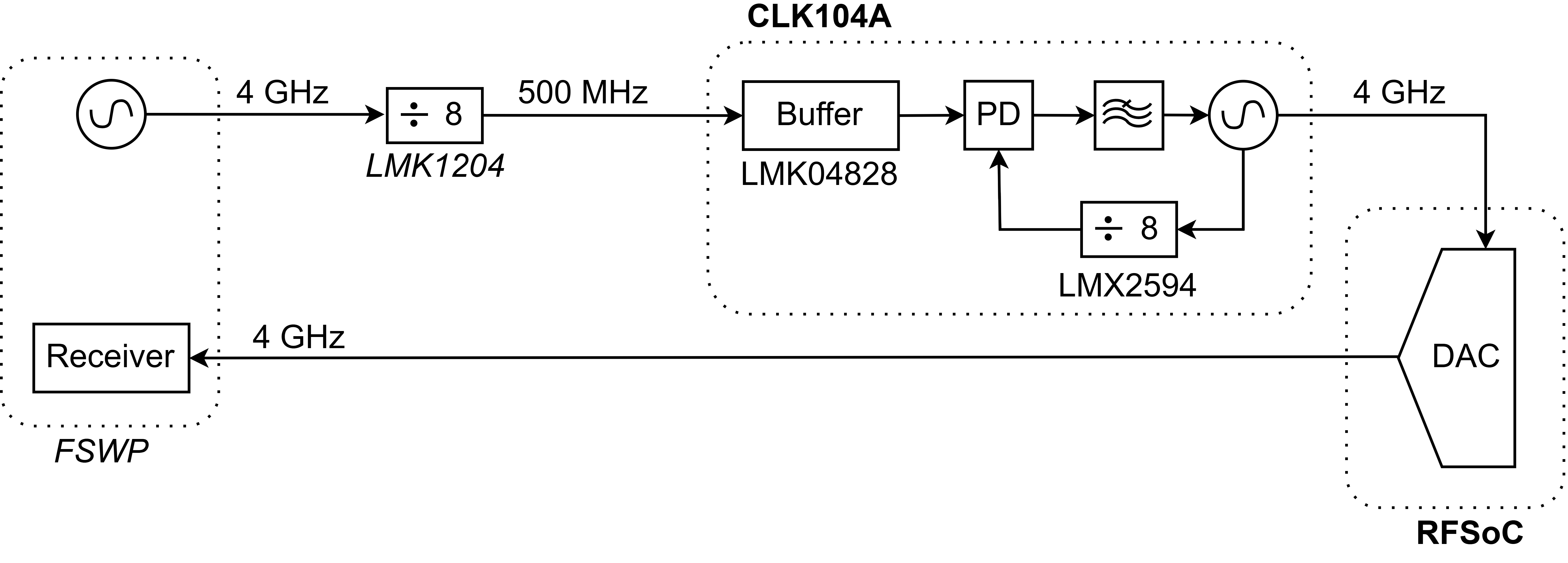}
    \caption{DAC + CLK104A Phase Noise Measurement Setup}
    \label{fig:placeholder}
\end{figure}

\begin{figure}[H]
    \centering
    \includegraphics[width=1\linewidth]{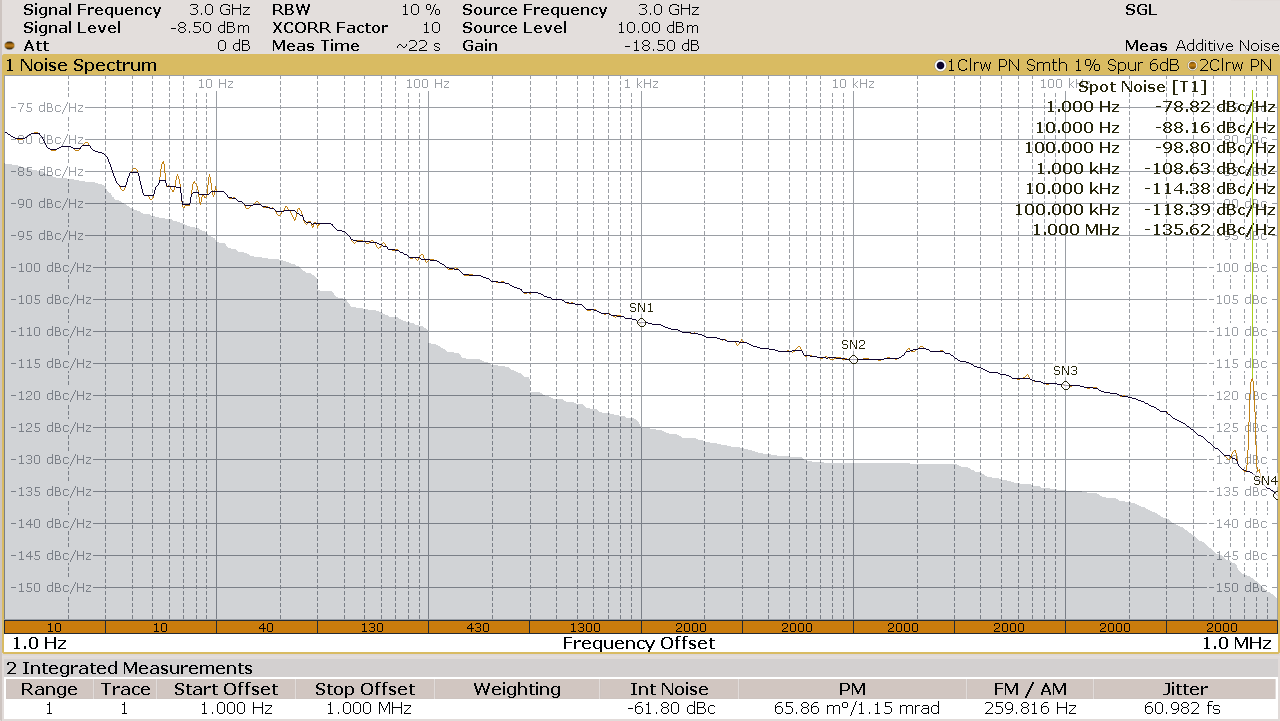}
    \caption{DAC + CLK104A 500MHz $RF_{\text{IN}}$, 4GS/s $f_s$, 3GHz $RF_{\text{OUT}}$} 60.9 fs [1 Hz, 1 MHz]
    \label{fig:placeholder}
\end{figure}

\begin{figure}[H]
    \centering
    \includegraphics[width=1\linewidth]{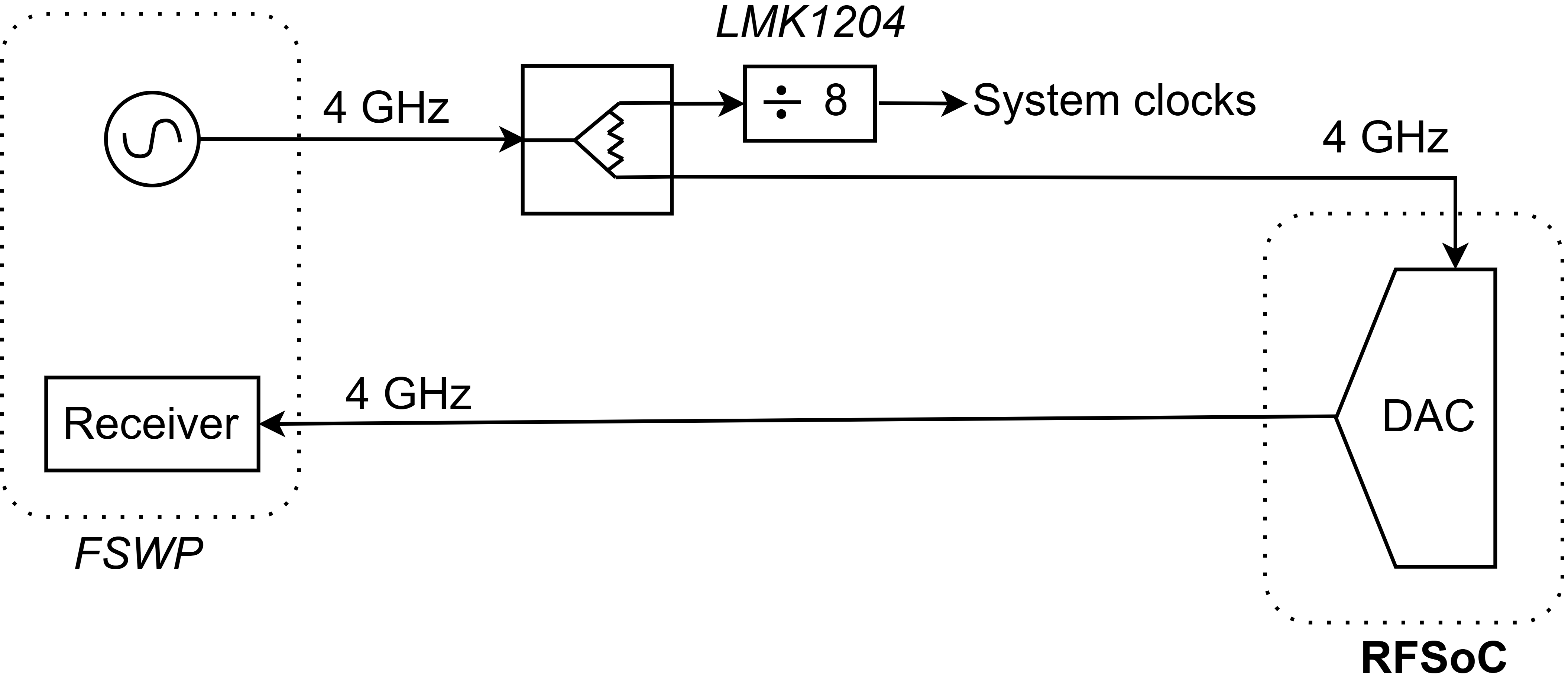}
    \caption{DAC Phase Noise Measurement Setup}
    \label{fig:placeholder}
\end{figure}

\begin{figure}[H]
    \centering
    \includegraphics[width=1\linewidth]{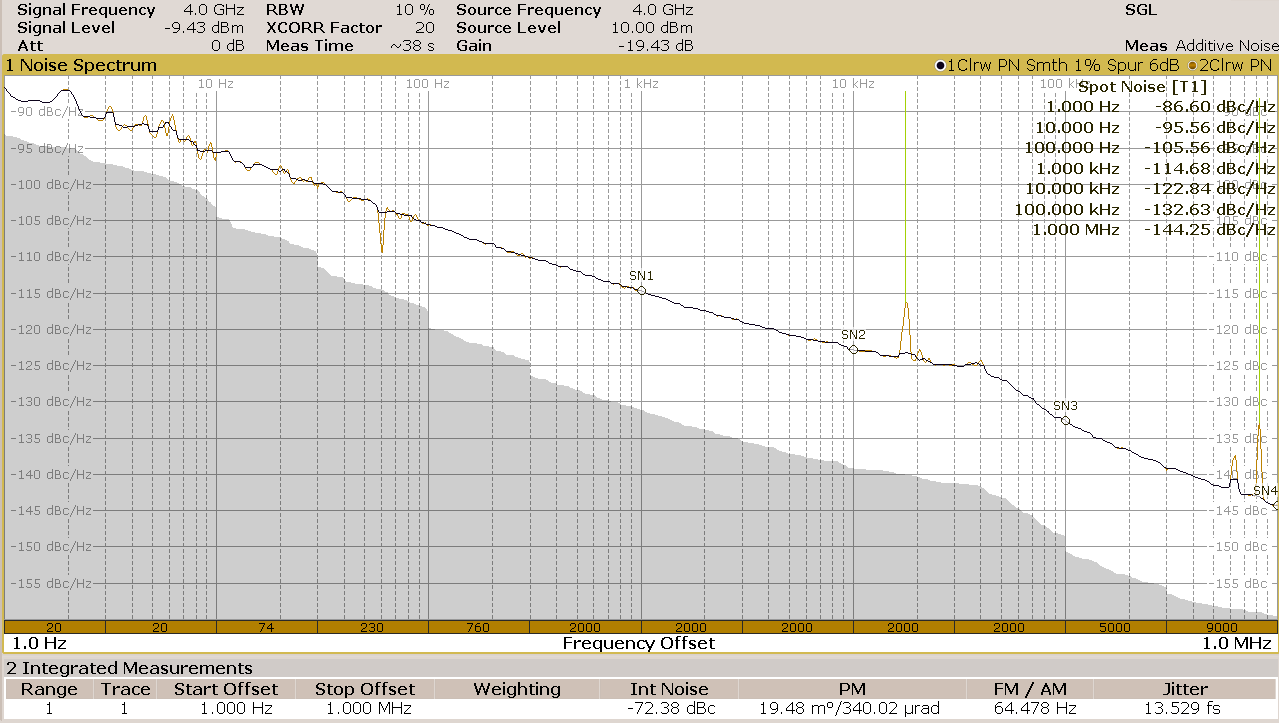}
    \caption{DAC Additive PN, 4GS/s $f_s$, 4GHz $RF_{\text{OUT}}$}  13.5 fs [1 Hz, 1 MHz]
    \label{fig:placeholder}
\end{figure}

\subsection*{ADC Phase Stability - Measurement Background}
ADC characterization is limited by the capture depth and rate. Current RFSoC firmware allows a capture depth of 32,768 samples (131.07$\mu$s with the 250 MHz DSP clock) and a maximum capture rate of \textasciitilde 37 Hz. Therefore these ADC characterization measurements of a CW source are pulsed performance results and must be characterized by the intrapulse and interpulse performance. 
\\
Intrapulse performance measures the stability of a single pulse, and can be evaluated as the mean of the standard deviations of the pulses.
$$\frac{1}{\tau_{\text{pulse}}} \le f \le \text{BW}_{\text{baseband}}$$
where $\tau_{\text{pulse}}$ is the pulse duration and $\text{BW}_{\text{baseband}}$ is the baseband bandwidth.

Interpulse measures the stability between pulses, and is characterized by the standard deviation of the pulse average values. Averaging within the pulse suppresses the intrapulse noise by the number of samples per capture. 

Averaging over a gate width $\tau_{\text{gate}}$ acts as a boxcar filter in the time domain, a sinc filter response in the frequency domain with a 3dB cutoff of $\approx \frac{0.443}{\tau_{\text{gate}}}$.
Frequency content up to $\frac{f_{\text{capture}}}{2}$ is directly measured and $\frac{f_{\text{capture}}}{2}$ to $\frac{0.443}{\tau_{\text{gate}}}$ is aliased into the measurement.
\\
Therefore the interpulse measurement describes the noise spectra of:
$$\frac{PRF}{N_{\text{captures}}} \le f \le \frac{0.443}{\tau_{\text{gate}}}$$
For these ALS capture measurement sets of 131.07$\mu$s pulses, 350 captures at 35 Hz, the interpulse jitter captures:
$0.1\text{Hz}\le f \le3.38\text{kHz}$
and the intrapulse jitter:
$7.63\text{kHz}\le f \le125\text{MHz}$
Future firmware upgrades will increase the capture rate by streaming to the DDR4, enabling complete characterization of the ADC phase noise across frequency offsets. 

\subsection*{ADC Phase Stability - Measurements}

The ADC measures $\varphi_{\text{ADC}} = \varphi_{f_s}-\varphi_{f_\text{RF}}$
, so the ADC and CLK104A noise can be characterized by using the same reference signal to generate the sampling clock and the RF input.
Since the CLK104A synthesizer PLL loop is \textasciitilde 250 kHz, the CLK104A added noise contributes primarily to the intrapulse noise, with a small broadband jitter increase from the buffer. 

The ADC noise is further evaluated by determining the uncorrelated noise between two channels fed the same RF signal, a technique used to characterize the fundamental receiver noise floor \cite{Larry:SRF2017_FRXBA02}:
\begin{figure}[H]
    \centering
    \includegraphics[width=1\linewidth]{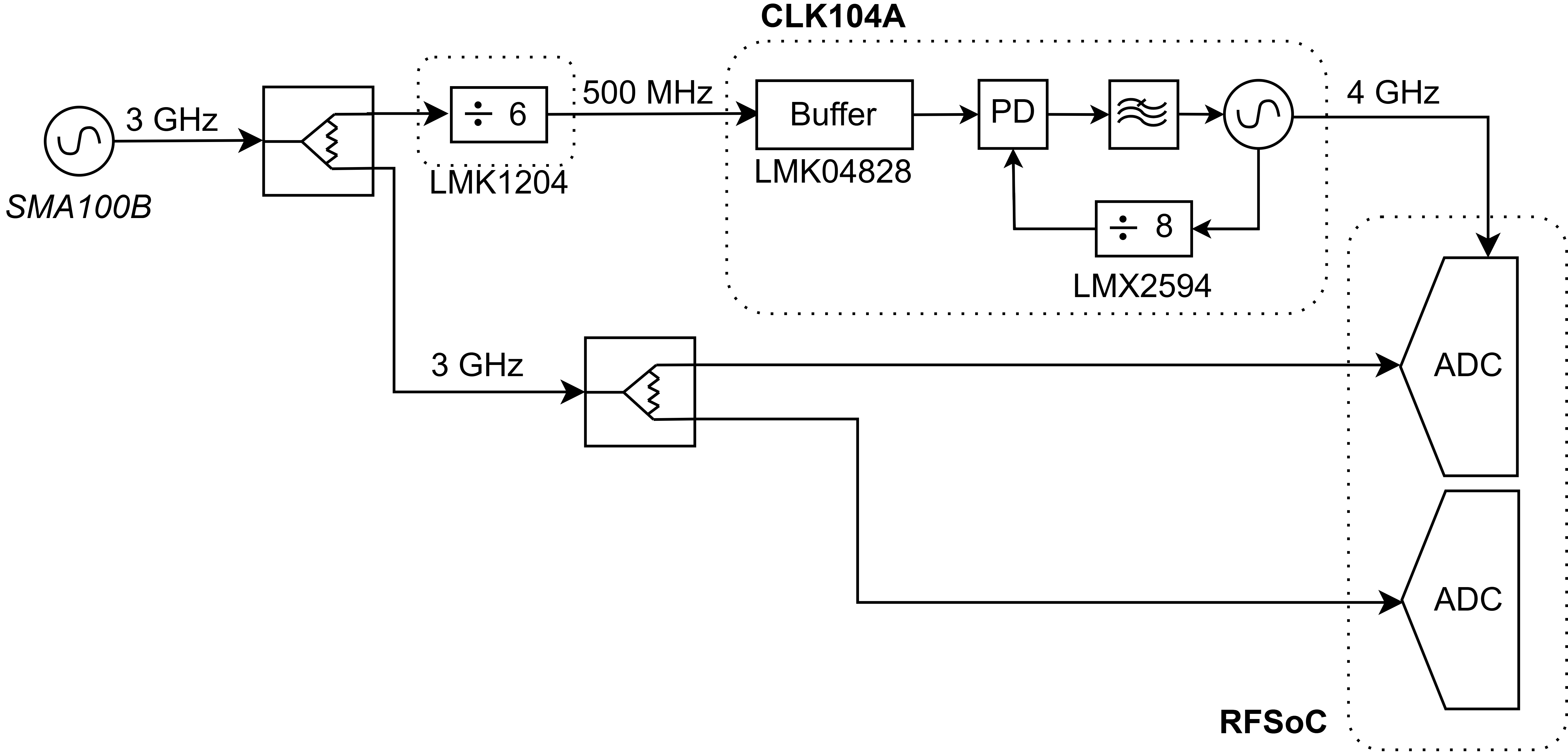}
    \caption{ADC Uncorrelated Noise Measurement Setup}
    \label{fig:placeholder}
\end{figure}

\begin{figure}[H]
    \centering
    \includegraphics[width=1\linewidth]{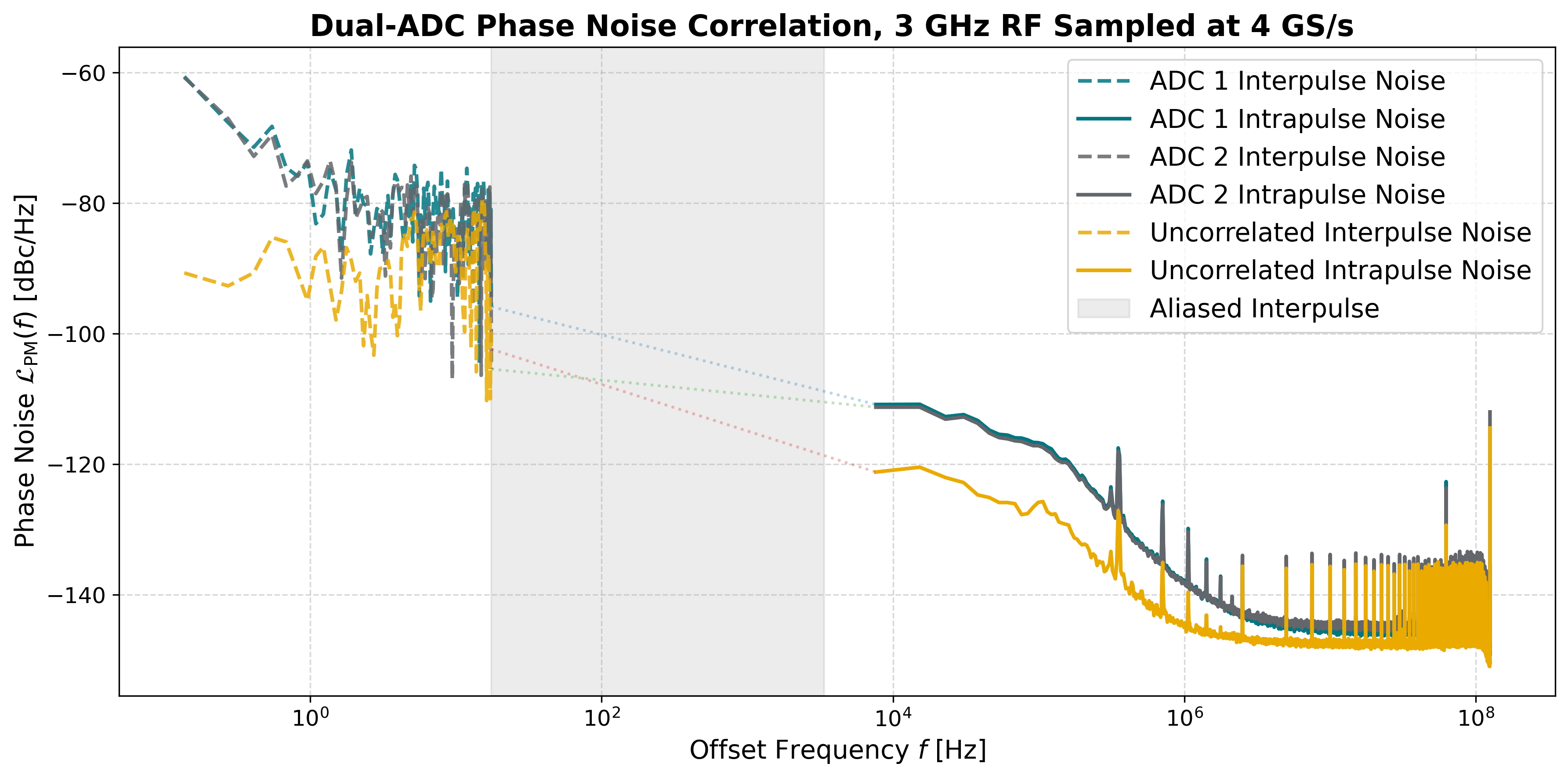}
    \caption{ADC Uncorrelated Noise}
    \label{fig:placeholder}
\end{figure}
The Pearson cross-correlation coefficient predicts the shared noise between the ADCs:

$$\rho_{12} = \frac{\text{Cov}(y_1, y_2)}{\sigma_1 \sigma_2} = \frac{\sigma_{\text{shared}}^2}{\sigma_1 \sigma_2}$$

And the variance of the normalized difference between the two ADCs predicts the ADC noise contribution:
$$\sigma_{\text{uncorr}} = \sqrt{\frac{\text{Var}(\text{ADC}_1 - \text{ADC}_2)}{2}}$$

\begin{table}[ht]
\centering
\begin{tabular}{l|r|r|r}
Parameter & Interpulse & Intrapulse & Total \\\hline
ADC 1 $\sigma_{\text{1}}$  & 44 fs & 73 fs & 85 fs \\
ADC 2 $\sigma_{\text{2}}$ & 41 fs & 76 fs & 86 fs \\

Correlation Coefficient ($\rho_{12}$) & +94\% & +67\% & --- \\
Residual ADC Noise ($\sigma_{\text{uncorr}}$) & 10 fs & 43 fs & 44 fs \\\hline
Sampling Clock Jitter ($\sigma_{\text{clk}}$) & --- & --- & 61 fs \\

Weighted ADC + CLK & 43 fs & 75 fs & 86 fs
\end{tabular}
\caption{\label{tab:dual_adc_bands}ADC Noise for 3 GHz Input, 4 GS/s $f_s$}
\end{table}

The alignment between the measured ADC + CLK104A phase noise and the correlation-weighted sum of the individual measurements provides some validation of the result.

\subsection*{Phase Noise Summary}

\begin{table}[ht]
\centering
\begin{tabular}{l|r|r}
Component & Jitter & Evaluation Freq\\\hline
CLK104A & 62 fs & 4 GHz\\
DAC + CLK104A &  61 fs & 4 GHz\\
DAC & 13 fs & 4 GHz\\
ADC + CLK104A & 86 fs & 3 GHz\\
ADC & 44 fs & 3 GHz
\end{tabular}
\caption{\label{tab:widgets}RFSoC Phase Noise Components}
\end{table}
The CLK104A and DAC were evaluated at 4 GHz, the current $f_s$, to enable the FSWP additive measurements. The ADC was evaluated at 3 GHz, the LINAC RF input.
\section*{Control Schemes}

\subsection*{PI Intrapulse Feedback}
The classic Low Level RF control system use Proportional-Integral (PI) feedback to track the amplitude and phase of the RF cavity voltage and adjust the RF drive signal to correct for errors \cite{simrock2022llrf}. 
The system latency limits the bandwidth and response speed of the PI loop.

\begin{table}[ht]
\centering
\begin{tabular}{l|r}
Delay Source & Value \\\hline
Phase lag due to delay & $360^\circ*BW_{loop}*latency$ \\
Phase PI integrator &  $90^\circ$ \\
Phase lag cavity & $atan(BW_{loop}/(BW_{cavity}/2))*180/\pi$

\end{tabular}
\caption{\label{tab:widgets}Phase Delays.}
\end{table}

The phase margin of the PI loop is $180^\circ-delay_{sum}$ and should be $> 45^\circ$ to ensure loop stability and maximize the loop bandwidth. For our estimated 450 ns total system latency (290 ns from the RFSoC \cite{du:2025}) and 0.4$\mu$ cavity fill time, this limits our maximum loop bandwidth to \textasciitilde100 kHz. 
The settling time for this loop is $ 3\text{-}4\tau_{\text{loop}}$, where
$$\tau_{\text{loop}} = \frac{1}{2\pi \cdot 100\text{ kHz}} \approx 1.6\,\mu\text{s}$$
Therefore this loop would require approximately $4.8\text{ to } 6.4\,\mu\text{s}$ to regulate and is not useful for this short pulse case.

\subsection*{Static Feedforward, and Adaptive Feedforward}
A combination of static feedforward (SFF) and adaptive feedforward (AFF) can reduce repetitive, synchronous errors in short pulse systems.

SFF is generated from a one-time calibration and is as the starting pulse shape applied by the arbitrary waveform generating (AWF) functionality of the DAC. This pulse shape is the inverse of the typical intrapulse error pattern of the pulse.

The AFF loop operates pulse-by-pulse to adjust the pulse shape as the error profile mutates due to slow changes. Stability requires a Q filter that performs intrapulse time averaging to limit the AFF's response to high-frequency noise: 
$$U_{n+1}(t) = U_n(t) + (q * E_n)(t)$$

Where $U_n$ is the drive waveform and $E_n$ is the error waveform. 
To ensure that the AFF only attempts to correct unaliased errors, the BW of the Q filter should be limited to frequency variations that fall below the pulse-to-pulse Nyquist limit (half the pulse repetition rate, $f_{\text{PRF}}/2$).

This limits the AFF correction feature size to 0.35/$BW_Q$. 
For ALS at 0.7 Hz : 1s, 
 limiting the AFF to DC corrections of the pulse average value.

\section*{Bounding Relevant Noise Sources}

\subsection*{Tx Noise}
Noise in the Tx chain will be injected into the cavity within its bandwidth and then considered as part of the error signal in a feedback system. With a PI loop, this sets the lower bound of relevant Tx noise to $PI_{\text{ENBW}}$. 
The upper bound is set by the ENBW of the RF cavity, $1/{2t_f}$.
For the ALS LINAC open-loop case, $$0 \text{ Hz} \le f \le \text{1.25 MHz}$$

The high-power RF system with a 1-2 MHz BW is sensitive to narrowband Tx spurs, and the normal conducting RF cavity applies \textasciitilde6 dB of suppression to the closest spurs. Through the direct modulation of the drive signal, AM-PM generation in the klystron, and IM3 products from the klystron, these spurs add \textasciitilde18 fs to the Tx error.

Noise in the Rx chain is an error source overlaid on the signal from the cavity, indistinguishable from true error.
In our DC-only interpulse scheme, the Rx noise is suppressed by averaging of the waveforms used for SFF calibration and intrapulse averaging of the feedforward DC error.

\section{Conclusion}
\label{sec:conclusion}

\begin{table}[ht]
\centering
\begin{tabular}{l|r|r}
Parameter & Phase Error & Timing Jitter \\\hline
DAC Phase Noise & 0.046$^\circ$ RMS & 43 fs RMS \\
TX Spurs & 0.020$^\circ$ RMS & 18 fs RMS \\\hline
Open-Loop Phase Error (RSS) & 0.050$^\circ$ RMS & 46 fs RMS \\
Diagnostic Phase Error (RSS) & 0.093$^\circ$ RMS & 86 fs RMS
\end{tabular}
\caption{\label{tab:open_loop_diag_summary}RFSoC LLRF Added Error Estimates}
\end{table}

For our open-loop controller case, the RFSoC added noise is 46 fs, 25$\%$ of our cavity field error budget. The 4 GS/s waveform capture currently achieves 86 fs, <0.1 $^\circ$ added phase error. 
In addition to the 3 GHz LINAC case described here, this controller will be used to upgrade the 125 MHz and 500 MHz Subharmonic Bunchers with longer RF pulses enabling intrapulse feedback \cite{du:2025}. Operational data from these installed systems is expected next year.

\section*{Acknowledgment}

This work is supported by the Office of Science, Office of Basic Energy Sciences, of the U.S.
Department of Energy under Contract No. DE-AC02-05CH11231.
Qiang Du led the LLRF RFSoC platform development with Shree Murthy, who optimized the system clocking architecture. Mike Chin designed the LBL208 and generously provided test hardware, ADC insights, and lab camaraderie. Our director Ben Flugstad drives us to improve the ALS through innovation, and our manager Angel Jurado leads the LINAC Upgrade project. 

\bibliographystyle{IEEEtran}
\bibliography{IEEEabrv,reference}

\end{document}